\documentclass[amssymb,prl,aps,amsmath,nofootinbib,twocolumn,preprintnumbers,showpacs]{revtex4}

\usepackage{color}
\usepackage{graphicx}
\usepackage[dvips]{hyperref}
\hypersetup{colorlinks,bookmarksopen,bookmarksnumbered,citecolor=blue,
linkcolor=black,pdfstartview=FitH,urlcolor=blue}
\usepackage{multirow}
\usepackage[utf8]{inputenc}
\usepackage{url}

\begin{document}

\title{Internal Bremsstrahlung Signature of Real Scalar Dark Matter\\
and Consistency with Thermal Relic Density}

\preprint{IPPP/13/54}
\preprint{DCPT/13/108}

\author{Takashi Toma}

\email{takashi.toma@durham.ac.uk}

\affiliation{Institute for Particle Physics Phenomenology, 
University of Durham, Durham, DH1 3LE, United Kingdom} 

\begin{abstract}
A gamma-ray excess from the galactic center consistent with line emission around $130~\mathrm{GeV}$ was recently found in the Fermi-LAT data.
Although the Fermi-LAT Collaboration has not confirmed its significance, such a signal would be a clear signature of Dark  Matter
 annihilation. Until now, there have been many attempts to explain the excess
 by Dark Matter. However these efforts tend to give too-small cross sections
 into photons if consistency with the correct thermal
 relic density
 of Dark Matter is required. In this letter, we consider a simple Yukawa interaction
 that can be compatible with both aspects and show which parameters
 are favored. 
\end{abstract}

\date{\today}


\maketitle

Observations of gamma-rays, cosmic ray positrons, anti-protons and
neutrinos are performing to look for Dark Matter (DM) signatures. In
particular, Fermi-LAT public data has recently been examined in detail, and
a gamma-ray excess around $130~\mathrm{GeV}$ from the region of the galactic center
 has been claimed~\cite{Bringmann:2012vr, Weniger:2012tx}. The
Fermi-LAT Collaboration also found the excess at
$135~\mathrm{GeV}$ independently~\cite{Bloom:2013mwa}, 
however
they found a much lower significance in the re-processed
data-set~\cite{Fermi-LAT:2013uma}. Many authors have 
provided models of this excess by monochromatic gamma-rays from DM
annihilation or decay, see for example~\cite{Tempel:2012ey,
Cline:2012nw, Choi:2012ap, Kyae:2012vi, Buckley:2012ws, Das:2012ys,
Kang:2012bq, Buchmuller:2012rc, Park:2012xq, Cline:2012bz, Wang:2012ts,
SchmidtHoberg:2012ip, Farzan:2012kk, Chalons:2012xf, Asano:2012zv,
Jackson:2013pjq, Kumar:2013ira, Ibarra:2013eda, Choi:2013eua}. 
If the source of the gamma-ray excess is DM annihilation, 
the required cross section into two photon is 
$\sigma{v}_{\gamma\gamma}=1.27\times10^{-27}~\mathrm{cm^3/s}$ 
for an Einasto DM density distribution; this value can change 
for a different DM profile~\cite{Weniger:2012tx}. 
The process of DM annihilation into two photon
is loop-suppressed because DM does not have electric charge. 
The loop-suppression factor is naively
expected to be $\alpha_{\mathrm{em}}^2/\left(4\pi\right)^2\sim10^{-7}$
compared with the annihilation cross section
$\sigma{v}_{\mathrm{th}}\sim10^{-26}~\mathrm{cm^3/s}$ where
$\alpha_{\mathrm{em}}$ is the electromagnetic fine structure constant. 
This value of $\sigma{v}_{\mathrm{th}}$ is needed to achieve the
correct relic density of DM. 
Thus, it seems difficult to be
consistent with the thermal relic
density of DM unless some enhancement of the cross section is
introduced~\cite{Tulin:2012uq}. 
In other words, if we assume DM is thermally produced, the gamma-ray
production cross section is fixed to a value that is too small to explain the
excess around $130~\mathrm{GeV}$. 

Another possibility is the explanation via Internal Bremsstrahlung (IB)
of Majorana DM~\cite{Bringmann:2012vr}. 
The possibility of explaining the $130~\mathrm{GeV}$
excess with IB has been explored in
refs.~\cite{Bringmann:2012vr, Bringmann:2012ez, Bergstrom:2012bd,
Shakya:2012fj, Garny:2013ama}. 
This process is the radiative
correction for the final state charged particles and the intermediate
particle. The IB process generates a line-like energy spectrum. 
The suppression factor compared with
$\sigma{v}_{\mathrm{th}}$ is roughly 
$\alpha_{\mathrm{em}}/\pi\sim10^{-3}$ which is larger than the
monochromatic photon case. 
Thus the IB process has better prospects than the monochromatic
$\gamma\gamma$ process from this point of view. 
However, even with the IB process, it seems difficult to be
compatible with the thermal relic density of DM. 
For standard p-wave annihilating neutralino DM, the IB signal is
still a factor of a few below the nominally required rate for the
observed density. 

In this letter, we consider IB for real scalar DM interacting with 
a fermionic mediator and a light fermion. As we
discuss below, the annihilation cross section into a light fermion-anti-fermion pair is expanded with the relative velocity of DM,
with a suppressed constant term. As a result, a higher order
term of the cross section can be dominant in the early universe, and the
cross section into gamma-rays becomes relatively large at present times, thus reconciling the relic density value and the interpretation of the
gamma-ray excess by DM annihilation. 


We consider a real scalar DM particle $\chi$ which has the following Yukawa
interaction with the electromagnetically charged fermion $f$ and the
fermionic mediator $\psi$
\begin{equation}
\mathcal{L}=y_L\chi\overline{\psi}P_Lf+\mathrm{h.c.},
\label{eq:int}
\end{equation}
where the fermion $f$ is typically a light lepton or a quark. The
annihilation cross section into $f\overline{f}$ is expanded as
$\sigma{v}_{f\overline{f}}=a+bv^2+cv^4+\mathcal{O}\left(v^6\right)$ with the
DM relative velocity $v$, and it is calculated under the approximation
of $m_f\ll m_\chi$ as 
\begin{eqnarray}
\sigma{v}_{f\overline{f}}\!\!&=&
\frac{y_L^4}{4\pi
m_\chi^2}\frac{m_f^2}{m_\chi^2}\frac{1}{\left(1+\mu\right)^2}
-\frac{y_L^4}{6\pi
m_\chi^2}\frac{m_f^2}{m_\chi^2}\frac{1+2\mu}{\left(1+\mu\right)^4}v^2
\nonumber\\
&&\!\!\!\!\!
+\frac{y_L^4}{60\pi m_\chi^2}\frac{1}{\left(1+\mu\right)^4}v^4
+\mathcal{O}\left(v^6\right),
\label{eq:2-body}
\end{eqnarray}
where the Yukawa coupling $y_L$ is assumed to be real, and the parameter $\mu$
is the ratio of masses defined as $\mu\equiv m_{\psi}^2/m_\chi^2>1$. 
The first and second terms of Eq.~(\ref{eq:2-body}),
which are called the s-wave and p-wave respectively, agree with the
appendix of ref.~\cite{Boehm:2003hm}. In addition, the d-wave term
which is proportional to $v^4$ is easily found to be the leading term in the limit of
$m_f\to0$. 
The s-wave and p-wave are suppressed by the factor $m_f^2/m_\chi^2$, thus
the d-wave becomes the 
dominant contribution to the cross section in the early universe 
when $\mu$ is not large enough. 
Conversely, the s-wave becomes dominant today even if the mass of 
particle $f$ is as low as the electron mass. 
The non-relativistic thermally averaged cross
section 
$\langle\sigma{v}_{f\overline{f}}\rangle$ 
is given by substituting $\langle{v^2}\rangle\to6T/m_\chi$ and
$\langle{v^4}\rangle\to60T^2/m_\chi^2$ where $T$ is the
temperature of the universe. 
This replacement coincides with refs.~\cite{Srednicki:1988ce, Gondolo:1990dk}. 
The thermally averaged cross section is
important to estimate the relic density of DM. 
The typical value of the temperature that sets the correct
relic density is roughly $m_\chi/T\approx20$-$25$. 

\begin{figure}[t]
\begin{center}
\includegraphics[scale=0.6]{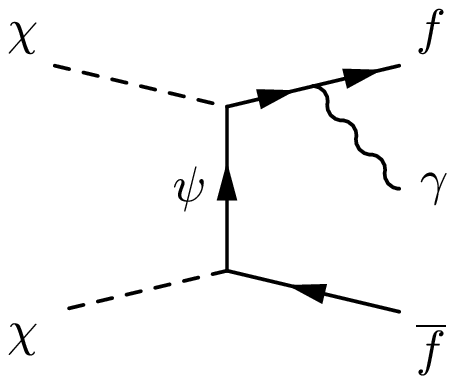}
\hspace{0.09cm}
\includegraphics[scale=0.6]{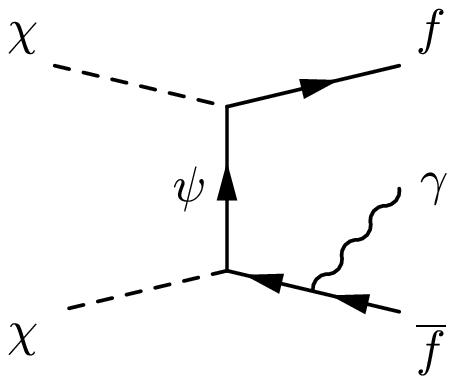}
\hspace{0.09cm}
\includegraphics[scale=0.6]{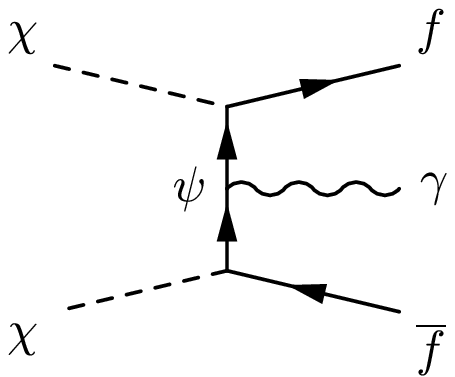}
\caption{Internal Bremsstrahlung processes of (real) scalar DM.}
\label{fig:vib}
\end{center}
\end{figure}

Next, we consider the radiative correction for the above 2-body process, 
that is $\chi\chi\to f\overline{f}\gamma$ shown in
Fig.~\ref{fig:vib}. This process is the IB of the real scalar DM, and the
emitted photon can be a gamma-ray signal which is 
comparable with the Fermi-LAT excess. 
The amplitude for the 
total IB is separated to two pieces of Final State
Radiation (FSR) and Virtual Internal Bremsstrahlung (VIB). 
One cannot generally treat the VIB diagram separately in a
gauge invariant manner, and it is only the sum of all three diagrams in
Fig.~\ref{fig:vib} that is gauge invariant.  
Here we define the FSR amplitude as the leading term of
the differential cross section in Eq.~(\ref{eq:fsr}),
and the VIB one as the amplitude removing the chiral suppression in
s-wave to the annihilation cross section. 
This definition of FSR and VIB makes the following discussion clear, 
however we note that it is different definition from
ref.~\cite{
Ciafaloni:2011sa}. 
Thus the differential cross section is expressed as
\begin{equation}
\frac{d\sigma{v}_{f\overline{f}\gamma}}{dx}=
\frac{d\sigma{v}_{f\overline{f}\gamma}^{\mathrm{FSR}}}{dx}
+\frac{d\sigma{v}_{f\overline{f}\gamma}^{\mathrm{VIB}}}{dx},
\label{eq:dcs}
\end{equation}
with $x=E_\gamma/m_\chi$, where 
the interference term between the FSR and VIB
amplitudes are neglected here. 

The first term in Eq.~(\ref{eq:dcs}) of FSR can be written in the model-independent way: 
\begin{equation}
\frac{d\sigma{v}_{f\overline{f}\gamma}^{\mathrm{FSR}}}{dx}\approx
\sigma{v}_{f\overline{f}}\frac{
Q^2\alpha_{\mathrm{em}}}{\pi}\frac{\left(1-x\right)^2+1}{x}
\log\biggl(\frac{4m_\chi^2\left(1-x\right)}{m_f^2}\biggr), 
\label{eq:fsr}
\end{equation}
where $Q$ stands for the electromagnetic charge of $\psi$ and $f$. 
A similar result is obtained for a bosonic final state, but
the $x$ dependence is different~\cite{Birkedal:2005ep}. 
The FSR differential cross section is proportional to the
2-body cross section $\sigma{v}_{f\overline{f}}$. 
This implies that if $m_f\ll m_\chi$, the FSR gives a very small
contribution and it can be negligible at present times. 
The energy spectrum of FSR is broad and it is not suitable to explain
the gamma-ray excess. 
If the FSR contribution is not suppressed, the energy spectrum of the
first term in Eq.~(\ref{eq:dcs}) invariably becomes greater than the
second term. 

The second term in Eq.~(\ref{eq:dcs}) represents the VIB
contribution~\cite{Bergstrom:1989jr, Flores:1989ru}. 
This process is well-known for enhancing the s-wave component in such chirally-suppressed models. 
The differential cross section of the VIB process for Majorana DM has
been calculated in ref.~\cite{Ciafaloni:2011sa, Bringmann:2007nk, Bergstrom:2008gr,
Asano:2011ik}. 
Similarly, it is calculated for real scalar DM by following
ref.~\cite{Ciafaloni:2011sa} as 
\begin{widetext}
\begin{equation}
\frac{d\sigma{v}_{f\overline{f}\gamma}^{\mathrm{VIB}}}{dx}=
\frac{Q^2\alpha_{\mathrm{em}}y_L^4}{4\pi^2m_\chi^2}\left(1-x\right)
\left[\frac{2x}{\left(\mu+1\right)\left(\mu+1-2x\right)}
-\frac{x}{\left(\mu+1-x\right)^2}
-\frac{\left(\mu+1\right)\left(\mu+1-2x\right)}{2\left(\mu+1-x\right)^3}
\log\left(\frac{\mu+1}{\mu+1-2x}\right)\right],
\label{eq:vib}
\end{equation}
and the total cross section is obtained by integrating
 Eq.~(\ref{eq:vib}) in the range of $0\leq x\lesssim 1$ as follows
\begin{equation}
\sigma{v}_{f\overline{f}\gamma}^{\mathrm{VIB}}=
\frac{Q^2\alpha_{\mathrm{em}}y_L^4}{8\pi^2m_\chi^2}
\left[\left(\mu+1\right)\left(\frac{\pi^2}{6}-\log^2\left(\frac{\mu+1}{2\mu}\right)
-2\mathrm{Li}_2\left(\frac{\mu+1}{2\mu}\right)\right)
+\frac{4\mu+3}{\mu+1}+\frac{4\mu^2-3\mu-1}{2\mu}\log\left(\frac{\mu-1}{\mu+1}
\right)\right],
\end{equation}
\end{widetext}
where $\mathrm{Li}_2(z)$ is the dilogarithm function defined by
$\mathrm{Li}_2(z)=-\int_0^1\log(1-zt)/tdt$. The above VIB cross
section for real scalar DM is a factor of $8$ times larger than that for Majorana DM. Note that the continuum gamma-ray spectrum due to
hadronization should be added in Eq.~(\ref{eq:dcs}) when the final state
particles are tauons or light quarks.

The above discussion is valid when the other interactions are sufficiently suppressed. Here we add the interaction with the right-handed
component of the fermion $f$, and we estimate how much hierarchy is
necessary among the interactions in order for the above scheme to work. If
the interaction with the left and right-handed fermion is 
\begin{equation}
\mathcal{L}=\chi\overline{\psi}\left(y_LP_L+y_RP_R\right)f+\mathrm{h.c.},
\end{equation}
the s-wave and p-wave components are not suppressed as in Eq.~(\ref{eq:2-body}), and
they are given by 
\begin{equation}
\sigma{v}_{f\overline{f}}=
\frac{y_L^2y_R^2}{\pi m_\chi^2}\frac{\mu}{\left(1+\mu\right)^2}
-\frac{y_L^2y_R^2}{3\pi
m_\chi^2}\frac{\mu+3\mu^2}{\left(1+\mu\right)^4}v^2
+\mathcal{O}\left(v^4\right).
\end{equation}
This formula coincides with the appendix of
ref.~\cite{Boehm:2003hm}. 
We can estimate the required condition among the parameters to
validate the above discussion. The s-wave component should be suppressed
enough compared with the d-wave, leading to the condition:
\begin{equation}
\left(\frac{y_R}{y_L}\right)^2\lesssim
\frac{v^4}{60\mu\left(1+\mu\right)^2}.
\label{eq:req}
\end{equation}
Therefore, $y_R/y_L\lesssim0.02$ is required when $\mu\sim1$
and $v^2\sim0.3$. 
For example even if $y_R$ does not exist at tree level, $y_R$ is
induced at one-loop level from the left-handed Yukawa coupling
$y_L$. When $\mu\sim1$, the right-handed Yukawa coupling $y_R$ is
then 
\begin{equation}
y_R\approx-\frac{y_L^3}{2(4\pi)^2}\mu\frac{m_f}{m_\chi}.
\end{equation}
This is sufficiently small compared with $y_L$ because of the factor
$m_f/m_\chi$, and the requirement Eq.~(\ref{eq:req}) is satisfied. 

We comment about  other DM models, such as complex scalar and fermionic DM. 
In the case of complex scalar DM, the s-wave component of the 2-body cross section is
suppressed by the factor $m_f^2/m_\chi^2$, just like real scalar DM, but
the p-wave component remains. 
Thus this framework discussed above does not work. 
For Majorana DM, the non-suppressed p-wave term is also present, and we cannot reconcile the DM relic abundance and
the explanation of the gamma-ray excess by DM annihilation. 
For Dirac DM case, the s-wave term exists for the 2-body process, and the FSR process
is always larger than the VIB process. Thus we cannot obtain
the line-like gamma-ray spectrum from VIB because the VIB signal is swamped
by the broad FSR spectrum.

We numerically analyze the consistency between the
thermal relic abundance of DM and the gamma-ray excess. 
In the following calculation, fermion $f$ is taken to be the
electron. 
%
The thermal relic density of DM is obtained by solving the Boltzmann
equation~\cite{Srednicki:1988ce, Gondolo:1990dk, Kolb:1989} 
\begin{equation}
\frac{z}{Y_{\mathrm{eq}}}\frac{dY}{dz}=-\frac{\Gamma}{H}
\left(\frac{Y^2}{Y_{\mathrm{eq}}^2}-1\right),
\label{eq:boltzmann}
\end{equation}
where $Y$ is defined as the DM number density $n_\chi$ divided by the
entropy density of the universe, and 
$Y_{\mathrm{eq}}$ is the value of $Y$ in thermal equilibrium. 
The reaction rate $\Gamma$ is defined as
$\Gamma\equiv\langle\sigma{v}\rangle n_{\chi}^{\mathrm{eq}}$ with the number
density in thermal
equilibrium, $H$ is the Hubble 
parameter and $z$ is a dimensionless parameter defined by $z=m_\chi/T$. 
The total cross section implies 
$\langle\sigma{v}\rangle=\langle\sigma{v}_{f\overline{f}}\rangle
+\langle\sigma{v}_{f\overline{f}\gamma}\rangle$. 
In general, we need approximately $\langle\sigma{v}\rangle\sim10^{-26}~\mathrm{cm^3/s}$ in order
to get the correct relic abundance of DM which is $\Omega_\chi
h^2=0.120$ observed by Planck~\cite{Ade:2013zuv}. 

\begin{figure}[t]
\begin{center}
\includegraphics[scale=0.8]{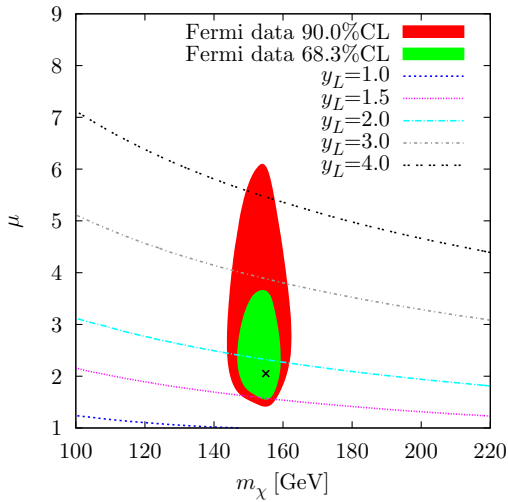}
\caption{The contours satisfying the DM relic density and 
 the favored region to fit to the gamma-ray excess in the
 $m_\chi$-$\mu$ plane. The fermion $f$ is assumed to be an
 electron here.}
\label{fig:result}
\end{center}
\end{figure}

We have only three parameters: $m_\chi$, $\mu$ and $y_L$. 
We solve the Boltzmann equation numerically with an implicit method. 
The contours of Yukawa coupling $y_L$ which satisfy the observed DM
relic density are depicted in Fig.~\ref{fig:result} in 
$m_\chi$-$\mu$ plane. We can see from the 
figure that a larger Yukawa coupling is required for larger $\mu$. 
Coannihilation between the DM $\chi$ and the mediator $\psi$ begins to
be effective in the region of $\mu\lesssim1.2$. 
For example, the process $\chi\psi\to fH$ can occur with the interaction
of Eq.~(\ref{eq:int}) and the SM Yukawa couplings where $H$ is the SM
Higgs. However, it would be small for the light SM charged particles. 
If the other interactions lead to effective coannihilation, 
this should be taken into account, as it may affect the
numerical analysis.

The gamma-ray flux coming from DM annihilation for the target region
$\Delta\Omega$ is given by 
\begin{eqnarray}
\frac{d\Phi_\gamma^{\mathrm{DM}}}{dE_\gamma}=
\frac{r_\odot}{8\pi}\frac{\rho_\odot^2}{m_\chi^2}\bar{J}
\langle\sigma{v}_{\gamma}\rangle
\frac{dN_\gamma}{dE_\gamma},
\label{eq:gamma-flux}
\end{eqnarray}
where 
$r_\odot=8.5~\mathrm{kpc}$ is the distance of the
earth from the galactic center, and $\rho_\odot=0.4~\mathrm{cm^3/s}$ is the
local DM density~\cite{Cirelli:2010xx}. The parameter $\bar{J}$ is defined as 
\begin{equation}
\bar{J}\equiv
\frac{1}{\Delta{\Omega}}\int dbd\ell\cos{b}\int_{\mathrm{l.o.s}}\frac{ds}{r_\odot}
\left(\frac{\rho(r,b,\ell)}{\rho_\odot}\right)^2,
\end{equation}
where $b$ and $\ell$ are the galactic latitude and longitude of the target region. 
The integral variable $s$ is related with the distance from the galactic
center $r$ as $r(s,b,\ell)=\sqrt{r_\odot^2+s^2-2r_\odot{s}\cos{b}\cos{\ell}}$. 
Note that the energy dependence arises
 only from the energy spectrum $dN_\gamma/dE_\gamma$, and
$\langle\sigma{v}_{\gamma}\rangle
dN_\gamma/dE_\gamma$ simply corresponds to 
Eq.~(\ref{eq:dcs}) at present situation. 
We use the generalized NFW profile~\cite{Navarro:1996gj} which is written
\begin{equation}
\rho(r)=\frac{\rho_s}{\left(r/r_s\right)^{\alpha}\left(1+r/r_s\right)^{3-\alpha}}. 
\end{equation}
It corresponds to the normal NFW profile if
$\alpha=1$. The parameter $\rho_s$ is the normalization factor in order to
fix to $\rho(r_\odot)=0.4~\mathrm{GeV/cm^3}$. The parameters $r_s$ and $\alpha$ are
taken as $r_s=20~\mathrm{kpc}$ and $\alpha=1.15$. 
This DM profile, as well as the parameter values, are the same as ref.~\cite{Weniger:2012ms} used to fit the excess around $130~\mathrm{GeV}$ in the Fermi-LAT data. 
We focus on the region of Reg4 in ref.~\cite{Weniger:2012tx,
Weniger:2012ms} to compare the gamma-ray flux from DM annihilation and
the claimed gamma-ray excess. 
The background of the gamma-ray flux is evaluated by the fitting
function~\cite{Weniger:2012tx} in the unit of
$\mathrm{GeV^{-1}cm^{-2}s^{-1}sr^{-1}}$ 
\begin{equation}
\frac{d\Phi_\gamma^{\mathrm{B}}}{dE_\gamma}=
2.4\times10^{-5} E_\gamma^{-2.55}. 
\end{equation}

\begin{figure}[t]
\begin{center}
\includegraphics[scale=0.7]{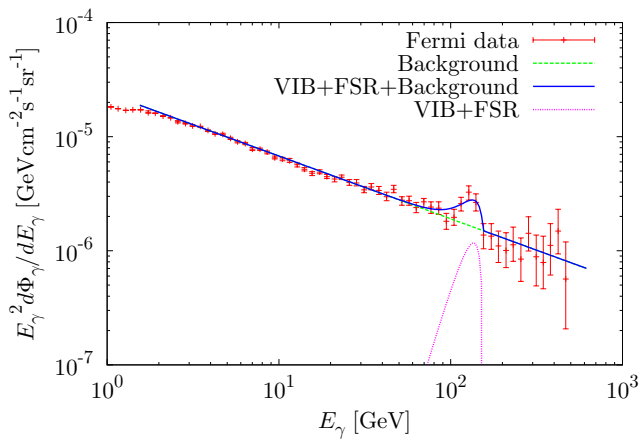}
\caption{Fitting to the gamma-ray excess via the VIB process. We use
 the best fit parameters found here. The data are taken from
 ref.~\cite{Weniger:2012ms}. Note that the energy
 dispersion of the Fermi instrument is included; it is
 approximately $10\%$ at $E_\gamma=100~\mathrm{GeV}$. This may
  alter the fit region.}
\label{fig:fitting}
\end{center}
\end{figure}

We find the best fit point in the parameter space of $m_\chi$, $\mu$ and
$y_L$ to give the gamma-ray excess. The 53 data points
counting from the the upper energy are taken from
ref.~\cite{Weniger:2012ms} and used for a chi-square analysis. 
Simultaneously, the constraint of the DM relic density is also imposed. 
As a result of the analysis, we get the best fit point of
$m_\chi=155~\mathrm{GeV}$, $\mu=2.05$ and 
$y_L=1.82$ with $\chi_{\mathrm{min}}^2=65.57$ ($51$ d.o.f). 
From the values, the cross section is calculated as
$\langle\sigma{v}_{f\overline{f}\gamma}\rangle=
4.72\times10^{-27}~\mathrm{cm^3/s}$, 
which is comparable with
$6.2\times10^{-27}~\mathrm{cm^3/s}$ obtained in
ref.~\cite{Bringmann:2012vr}, 
while parameter setting is slightly different.
The favored $m_\chi$-$\mu$ region to fit to the gamma-ray excess is
shown in Fig.~\ref{fig:result} 
where the Yukawa coupling $y_L$ is fixed by the constraint of thermal
relic density of DM at each point. 
From the figure, we can see that the DM abundance and the gamma-ray excess
coming from DM annihilation are consistent each other. 
The favored region in large $\mu$ would be slightly
changed if the monochromatic photon induced by the box diagrams is taken into
account in the model~\cite{Giacchino:2013bta}. 
The fitting of the gamma-ray excess with the evaluated values is
depicted in Fig.~\ref{fig:fitting}.

The Yukawa interaction considered here contributes to the anomalous
magnetic moment of fermion $f$, thus it may constrain the strength of
the interaction. 
The anomalous magnetic moment of $f$ is
calculated from the Yukawa interaction Eq.~(\ref{eq:int}) as~\cite{Boehm:2003hm}
\begin{equation}
\delta{a}_f=\frac{y_L^2}{(4\pi)^2}\frac{m_f^2}{m_\chi^2}
\frac{2+3\mu-6\mu^2+\mu^3+6\mu\log{\mu}}{6(1-\mu)^4}.
\end{equation}
The current experimental bound~\cite{Hanneke:2008tm, Hanneke:2010au} for
the electron anomalous magnetic 
moment is given as $\delta{a}_{e}\equiv
a_{e}(\mathrm{SM})-a_{e}(\mathrm{exp})=1.06\times10^{-12}$~\cite{Aoyama:2012wj},
while the value at the fitting point is two orders of magnitude lower:
$\delta{a}_e\approx9.4\times10^{-15}$.
For the muon, the experimental bound is
$\delta{a}_{\mu}=25.5\times10^{-10}$~\cite{Bennett:2006fi, vonWeitershausen:2010zr}
and our value is $4.0\times10^{-10}$, one order of magnitude below the bound. However, for example if we have simultaneous Yukawa couplings to both the electron and muon, the Yukawa coupling is extremely constrained by charged
lepton flavor violating processes like $\mu\to e\gamma$ unless destructive interference occurs.

We have discussed scalar DM having Yukawa interaction with the 
left-handed light fermion $f$ and the mediator $\psi$. The annihilation cross
section of the DM into $f\overline{f}$ is highly suppressed since the
s-wave and p-wave are proportional to the ratio of masses
$m_f^2/m_\chi^2$. As a result, the d-wave can be dominant  
in the early universe, and the DM relic abundance is obtained
by the d-wave cross section. 
Simultaneously, the VIB component of the radiative correction for the
process that is $\chi\chi\to f\overline{f}\gamma$ has s-wave and it
gives the line-like gamma-ray signal. 
The recently-observed gamma-ray excess is well explained without
inconsistency with the thermal relic density of DM. 
We have three parameters of $m_\chi$, $\mu$ and $y_L$, and 
obtained the best fit point $m_\chi=155~\mathrm{GeV}$, $\mu=2.05$ and $y_L=1.82$ 
by fitting to the gamma-ray excess with the constraint
of the DM thermal relic density. 

The framework discussed here works when the other interactions of DM
are small enough. The lightest right-handed sneutrino DM in
supersymmetric extended models would be a realistic candidate since
the chargino plays a role in the mediator $\psi$. 
To do that, the neutrino Yukawa interaction should be large. 
Therefore supersymmetric radiative seesaw models such as
refs.~\cite{Suematsu:2010nd, Fukuoka:2010kx} would be 
promising concrete models to implement this scheme because the neutrino
Yukawa coupling can be order one and tiny neutrino masses are derived
without contradiction. 
In addition, inverse seesaw models with supersymmetry
would also be good candidates for the framework~\cite{Khalil:2011tb}.

\begin{acknowledgments}
The author would like to thank C\'eline B\oe hm, Silvia Pascoli and
Thomas Schwetz for useful discussion, and Aaron Vincent for careful
 reading of the manuscript and valuable comments. 
The author acknowledges support from the European ITN project
(FP7-PEOPLE-2011-ITN, PITN-GA-2011-289442-INVISIBLES). 
Numerical computation in this work was partially carried out at the
Yukawa Institute Computer Facility.

{\bf Note added :}
An article finding similar results was almost simultaneously 
 published on the arXiv~\cite{Giacchino:2013bta}. 
\end{acknowledgments}

\end{document}